# Cosmogenic $^7$Be in ground level air in Rostov-on-Don (Russia) (2001-2011)

E.A. Buraeva [a], V.S. Malyshevsky [a,1], V.C. Nephedov [a], B.I. Shramenko [b], V.V. Stasov [a], L.V. Zorina [a]

[a] *Southern Federal University, 344090, Rostov-on-Don, Russia*
[b] *National Science Center «Kharkov Physical-Technical Institute», 61108, Kharkov, Ukraine*

**Abstract**

The deposition flux of cosmogenic $^7$Be in the industrial city Rostov-on-Don (Russia) from 2001 to 2011 has been measured. The variations of annual $^7$Be deposition flux appear to be mainly correlated with the number of the meteorological parameters and solar activity. For the first time correlations of the volume activity of cosmogenic $^7$Be with such meteorological parameters as temperature, precipitation, wind speed, atmospheric pressure, relative humidity are identified.

## 1. Introduction

To date, the monitoring of radionuclides in the atmospheric boundary layer suggests that a substantial contribution to the radioactivity of the surface air gives a short-lived isotope $^7$Be ($T_{1/2}$ = 52.3 days) of the cosmogenic origin. Variations of the contents of $^7$Be in the air are associated with the solar activity and have a characteristic seasonal variation and latitude dependence. Due to the rapid decay, its activity in plants varies depending on the synoptic conditions. Therefore, $^7$Be is of interest not only in terms of radiation exposure on biological systems but can also be an indicator of rates of exchange in plants, and, as a consequence, an indicator of the accumulation by natural environments of pollutants entering the atmosphere. $^7$Be is one of the few radionuclides that are independent of anthropogenic activity. Therefore it can be used as a monitor to detect sources of other radionuclides into the atmosphere. This is what makes $^7$Be a convenient indicator for a rapid assessment of potential air pollution and air exchange in the environment. Therefore, the study of occurrence processes, transportation and migration of the radionuclide $^7$Be in the environment is of great interest.

It is believed that the main reactions leading to the formation of beryllium isotopes in Earth's atmosphere occur in the interaction of cosmic rays with nuclei of nitrogen and oxygen (Nagai, et.al., 2000), which are the main components of the air. These are the so-called

---
[1] Corresponding author. E-mail address: vsmalyshevsky@sfedu.ru

"spallation" reactions: $^{14}$N(p,X)$^7$Be, $^{16}$O(p,X)$^7$Be (up to 70–80%), $^{14}$N(n,X)$^7$Be and $^{16}$O(n,X)$^7$Be (up to 20–30%) (Yoshimori, 2005). Another possible mechanism of formation of beryllium isotope $^7$Be in the upper atmosphere can be the photo-nuclear reactions $^{14}$N(γ,X)$^7$Be, $^{16}$O(γ,X)$^7$Be and $^{12}$C(γ,X)$^7$Be. It is shown (Bezuglov, et.al., 2012) that the contribution of the photonuclear mechanism is comparable with the contribution of the proton and neutron channel of $^7$Be formation in the atmosphere. The contribution of the photonuclear reactions to the total $^7$Be production in the atmosphere is not less than 10%.

During changes in the solar activity (reduced number of sun spots is the Wolf number, URL) within the 11-year solar cycle and aperiodic bursts of solar activity, the geomagnetic field changes, cosmic rays are deflected and, correspondingly, the $^7$Be production rate changes (Taplos, et.al., 2005). A decrease of the $^7$Be production rate corresponds to an increase of solar activity (increase of the Wolf number) and vice versa, i.e., there is an anticorrelation between the $^7$Be content in the atmospheric air and the Wolf number with coefficient $k = −0.81$ according to Ioannidou A. (2005) and $k = −0.83 ± 0.03$ according to Stozhkov (2002). Over the 11-year solar cycle, the yearly average content at the maximum and minimum differs by approximately 45%. The $^7$Be production rate also depends on the geographical coordinates of the observations station because of the effect of the Earth's magnetic field on the cosmic ray distribution (Papastefanou, et.al. 2004).

Long (more than two cycles of solar activity) systematic measurements on the global network of stations must be performed in order to determine reliably the relation between the $^7$Be volume activity in the air layer at the ground and the solar activity against the background of variations of a different origin. The results of the determination of $^7$Be in the atmosphere in 1974–1999 at 26 stations were analyzed by Taplos, et.al., (2005). The existence of the anticorrelation indicated above, which explains about 54% of all temporal variations of the $^7$Be for stations in Australia, New Zealand, and North America and only 18% of the variations for the stations in South America and Antarctica, has been proven.

Long-time measurements (1987–2003) were performed by Papastefanou, et.al. (2004) at temperate latitudes (40°38′). Under especially favorable conditions (regularity of measurements of the meteorological parameters, absence of any effect due to some of them, and so forth), a correlation between the $^7$Be content and the Wolf number can be determined reliably. Thus, measurements performed under the conditions of a dry and hot climate showed (Al-Azmi, et.al., 2001) that the changes of the yearly average volume activity of $^7$Be depend on the Wolf number. A correlation cannot be established under different, less favorable, conditions (Petrova, et.al. 2003).

Almost immediately after they are formed, the $^7$Be nuclei precipitate in submicron-size aerosols, and transport with air masses, settling, and washing out by precipitation determines their subsequent fate. The methods used to determine the life time of aerosols (the period of time during which half of the initial content of the aerosols is removed from the atmosphere) and the results obtained are presented by Papastefanou (2006) together with data for other observation points (Greece, Germany, California, Hong Kong). It has been suggested that the data be divided into two groups: 2.6–15 days (average 8.8 days) for the air layer at the ground and 21–35.4 days (average 28.2 days) for the troposphere. According to other ideas, the first group describes tropospheric and the second stratospheric aerosols. Estimates obtained using the model of Koch, et.al., (1996) give 24–30 days for tropospheric aerosols and 1 yr. for stratospheric aerosols.

## 2. Materials and methods

The variations of the $^7$Be volume activity in the air layer at the ground depend on the exchange of air masses between the stratospheric and tropospheric reservoirs, dry and wet fallout, and tropospheric processes (vertical transport, advection) [3]. Measurements of the $^7$Be content in aerosols (1 per week) and precipitation (1per month) are performed at the aspiration station of the Southern Federal University (Rostov-on-Don, Russia) in 2001–20011 as part of the monitoring of the radioactivity of the atmospheric layer at the ground in Rostov-on-Don (47°14′ NL; 39°42′ EL). The location of the station at temperate latitudes with a temperate continental climate and comparatively low precipitation imparts special significance to the systematic monitoring of 7Be in the atmosphere.

A ventilation setup with a filter consisting of FPP-15-1.7 Petryanov fabric with total area 0.56 m$^2$ and a liquid Lambrecht micromanometer were used to obtain the samples. According to the measurements, the air flow rates were approximately 630 m$^3$/h initially ("fresh" filter) and 510 m$^3$/h after 7 days of exposure. The exposed filter was air dried and pressed into 35 mm in diameter and 10–30 mm high pellets. Three or four days after the filter was removed, the γ-ray spectrum was measured in 12–24 h with a Ge(Li) or HPGe detector of the low-background setup. $^7$Be was determined according to the 477 keV peak. The dust content in air was found according to the mass difference between the exposed and clean filter.

## 3. Results and discussion

As a result of continuous measurements of $^7$Be volume activity in surface air of Rostov-on-Don for the period 2001-2011 we found that the atmospheric aerosols concentration varies from 0.025 to 27.0 mBq/m$^3$ mBq/m3, with an average grade on record 6,0 mBq/m$^3$.

Analysis showed that the data are sufficient to determine the anticorrelation between the $^7$Be volume activity and the solar activity. We were able to establish the dependence of the $^7$Be volume activity for the second half of the 23rd and the first half of the 24th cycles of solar activity. The yearly and monthly average $^7$Be volume activity (Figs. 1 and 2, respectively) increases toward the end of the 23rd cycle and reaches a maximum at solar activity minimum of the 24th cycle at 2008. The corresponding scatter plot for the period 2001-2011 is shown in Fig. 4f. There is an anticorrelation between the $^7$Be content in the atmospheric air and the Wolf number with coefficient $k = -0.42$.

It is clear the volume activity of $^7$Be does not react to short-time variations of the Wolf numbers. To reliably determine the dependence of the $^7$Be concentration in the surface layer of the atmosphere on solar activity variations the systematic determination of 7Be in the global network of stations (more than 2 cycles of solar activity) are needed.

The seasonal variation of $^7$Be in aerosols, which is well known for different latitudes and climatic conditions and is associated with the spring rearrangement of the atmosphere in the stratosphere–troposphere system, is quite clearly detected. As a rule, the seasonal variation of the $^7$Be volume activity exhibits a spring–summer maximum and an autumn–winter minimum. Thus, for temperature latitudes (Greece) the summer maximum is 7.29–6.96 mBq/m$^3$ and the winter minimum is 2.75–4.09 mBq/m$^3$ (Ioannidou, et.al., 2005). For Moscow (Russia), the spring–summer maximum is 4.3–4.6 mBq/m$^3$ and the autumn–winter minimum is 2.6–3.3 mBq/m$^3$ (Petrova, et.al. 2003). Our data show that spring–summer maximum of the $^7$Be volume activity in aerosols is observed yearly (Fig. 2) and on the average over 2001-2011 (Fig. 3). The average ratios of the maximum to minimum values of the seasonal average of the $^7$Be content equal approximately 2.6 (for Moscow 1.6 over 1996–2001). Fourier analysis of the entire set of data over ten years confirms the seasonal variation – the period of the first dominant harmonic is 52 weeks. Previously obtained (Buraeva, et.al., 2007) the five-year average (2001–2005) of $^7$Be volume activity in aerosols in Rostov-on-Don was about 3.9 mBq/m$^3$. The increase in the average $^7$Be volume activity nearly doubled (up to 6.0 mBq/m$^3$) due to the minimum of solar activity in 2008 and growth in the production of cosmogenic beryllium in the atmosphere as a consequence.

The salient features of the seasonal variation of the $^7$Be content in aerosols from one year to another are related with the changes in the meteorological conditions (temperature, precipitation, wind speed, atmospheric pressure, relative humidity). The Table 1 shows the average meteorological parameters for the city of Rostov-on-Don during the observation period 2001-2011. The corresponding scatter plot for the period 2001-2011 and correlation coefficients are shown in Fig. 4.

At temperate latitudes, the amount of precipitation has the greate effect on the $^7$Be concentration. The generalized results of an analysis of the relation between the $^7$Be content in aerosols and precipitation show (Fig. 4c) the presence of anticorrelation with a coefficient $k = -0.21$. The opposite relation between the $^7$Be content in aerosols and precipitation is due to selective washing out of the atmosphere by precipitation. The volume activity and atmospheric pressure are similarly linked (Fig. 4a). This is understandable, because the increase in precipitation is accompanied by a decrease in atmospheric pressure.

After falling onto the ground, $^7$Be accumulates in the soil–vegetation cover. The maximum amount of precipitation, occurring in June–July, decreases the $^7$Be concentration in aerosols immediately after its summer maximum in July. On the whole, this is in agreement with the data on the effect of precipitation on the $^7$Be content in the atmosphere at temperate latitudes (Ioannidou, et.al., 2005). Wet precipitation is the most effective mechanism for removing $^7$Be from the atmosphere. The wash-out coefficient is estimated by Buraeva, et.al., (2007) to be 30–60% and depends on the dispersity of the aerosol and the type of precipitation (snow, rain, downpour, protracted), which lower the $^7$Be content almost all year.

Seasonal course of $^7$Be repeats the change in temperature for the observation period. There is a direct correlation of these parameters. The temperature dependence of the $^7$Be concentration (Fig. 4d) is determined over a period of ten years and the correlation with temperature (Fig. 4, $k = 0.61$) has been established.

During the observation period the wind speed since 2006 has doubled (see Table 1), which may assist to increase the role of wind lifting of radionuclides in the near-surface atmosphere. Since 2006 the east and north-east winds was dominated, and in 2009 just east wind was dominated. These changes in meteorological parameters (mostly wind speed/direction, and relative humidity) contribute to the winds lifting of the soil dust and increase the volume activity of beryllium in the air. The correlation with wind speed is $k = 0.34$ (Fig. 4e).

The dependence of the activity concentration of $^7$Be in the surface atmosphere of Rostov-on-Don on the relative humidity of the air is inverted (Fig. 4b, $k = -0.62$). Basically, at the highs of $^7$Be activity in surface air the relative humidity decreases. In most cases the maximum volume activity of $^7$Be for the period of July to September with the lowest values of relative humidity (below 50%). Such a dependence of the $^7$Be in atmospheric aerosols on the amount of rainfall and relative humidity confirms the washout of aerosols by the precipitation.

4. Conclusions

On the whole, the results of our analysis of the $^7$Be content in atmospheric aerosols and meteorological parameters illustrate the main features of the variation of these quantities and their relation with the regional climatic characteristics. The variations of annual $^7$Be deposition flux appear to be mainly correlated with the number of the meteorological parameters and solar activity. For the first time correlations of the volume activity of cosmogenic $^7$Be with such meteorological parameters as temperature, precipitation, wind speed, atmospheric pressure, relative humidity are identified. The correlations allow one to predict radioactive contamination of the atmosphere in future.

**Acknowledgement**

This work was supported by Federal Program of the Russian Ministry of Science and Education "Scientific and scientific-pedagogical personnel of innovative Russia" (grant number 14.A18.21.0633). The final stage of this work was supported by Russian Foundation for Basic Research and by National Academy of Sciences of Ukraine grant number 12-08-90401-Ukr_a.

**Table 1.** Averaged meteorological parameters

| Year | Rainfall, mm | Temperature, $^0$C | Wind speed, m/s | Relative humidity, % | Atmospheric pressure, mm Hg |
|---|---|---|---|---|---|
| 2001 | 67,0 | 10,0 | 1,9 | 71,7 | 755,5 |
| 2002 | 46,0 | 10,3 | 1,8 | 69,2 | 756,0 |
| 2003 | 52,0 | 9,0 | 1,9 | 70,8 | 756,8 |
| 2004 | 78,0 | 10,2 | 1,7 | 75,1 | 755,3 |
| 2005 | 58,0 | 10,8 | 1,9 | 71,6 | 756,2 |
| 2006 | 47,0 | 10,4 | 3,1 | 69,8 | 755,9 |
| 2007 | 31,0 | 12,3 | 4,2 | 64,8 | 755,3 |
| 2008 | 36,0 | 10,8 | 4,3 | 69,5 | 756,3 |
| 2009 | 51,0 | 10,9 | 4,4 | 70,8 | 755,5 |
| 2010 | 45,0 | 11,9 | 5,1 | 68,8 | 755,0 |
| 2011 | 48,0 | 10,1 | 4,4 | 69,2 | 756,1 |

# Captions

**Fig. 1**. The yearly average $^7$Be volume activity and solar activity for period 2001-2011 (note that to be placed on the same graph the $^7$Be volume activity increased 10 times).

**Fig. 2.** The monthly average $^7$Be volume activity and solar activity for period 2001-2011 (note that to be placed on the same graph the $^7$Be volume activity increased 10 times)..

**Fig. 3.** The yearly average for period 2001-2011 seasonal variation of $^7$Be volume activity and solar activity (note that to be placed on the same graph the $^7$Be volume activity increased 10 times).

**Fig. 4**. Scatterplot showing the correlation between meteorological parameters, solar activity and volume activity.

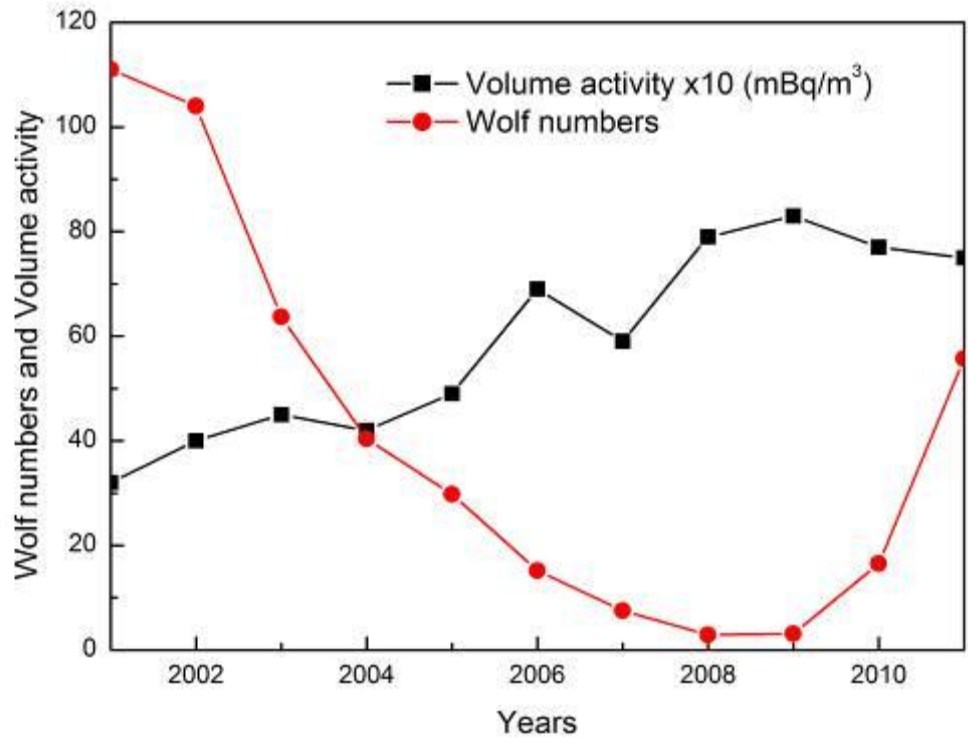

Fig. 1

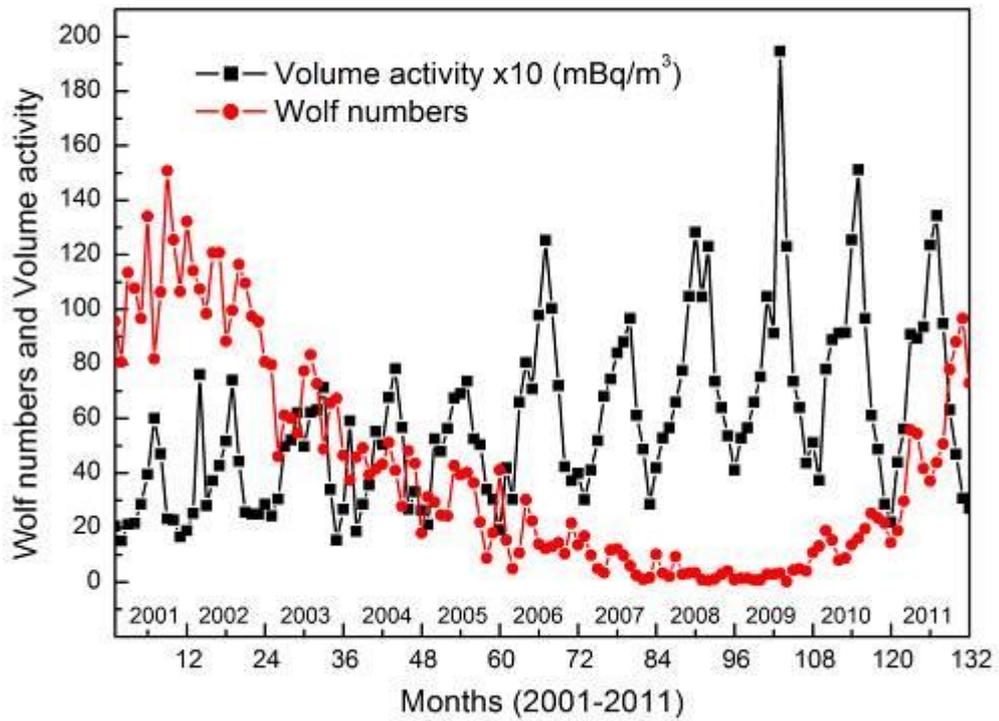

Fig. 2

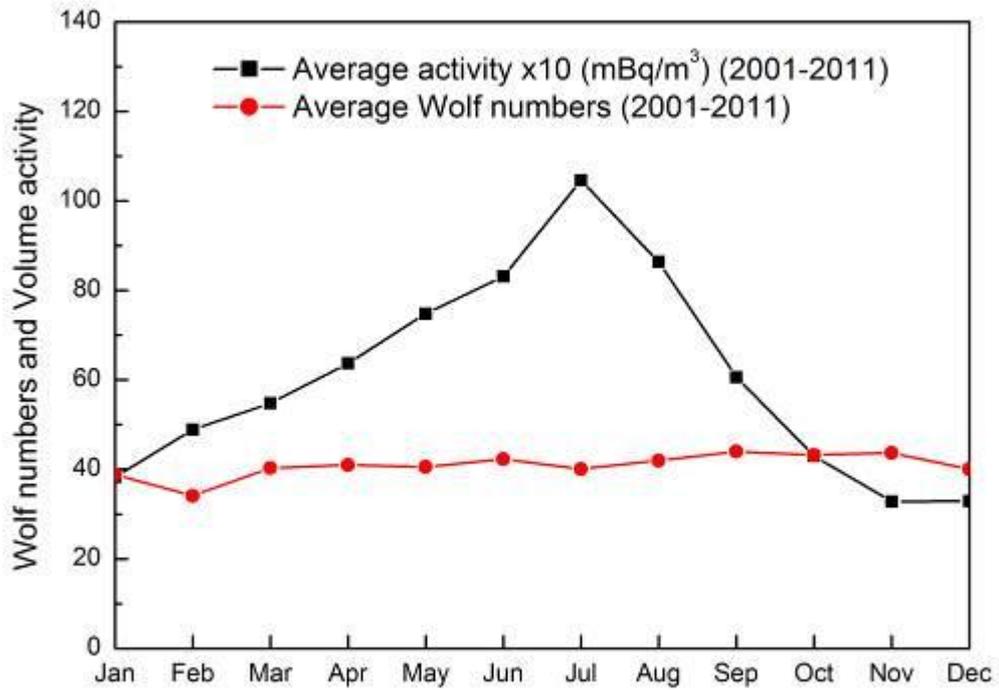

Fig. 3

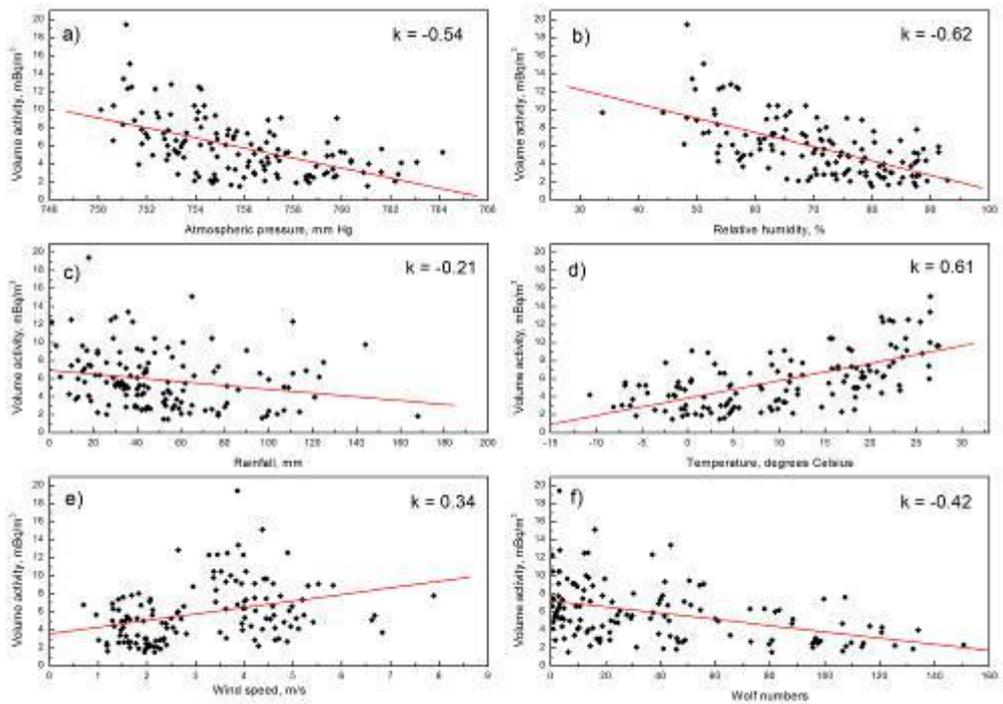

Fig. 4